\documentclass{IEEEcsmag}

\usepackage[colorlinks,urlcolor=blue,linkcolor=blue,citecolor=blue]{hyperref}
\expandafter\def\expandafter\UrlBreaks\expandafter{\UrlBreaks\do\/\do\*\do\-\do\~\do\'\do\"\do\-}
\usepackage{upmath,color}

\jvol{XX}
\jnum{XX}
\paper{X}
\jmonth{November}
\jname{IEEE Software}
\jtitle{Special Issue on the Wellbeing for Resilience: Developers Thrive}
\pubyear{2023}

\usepackage{xcolor}

\usepackage{tabularx}
\usepackage{colortbl}

\usepackage{url}
\usepackage{hyperref}

\usepackage{graphicx}
\usepackage{array}			
\usepackage{multirow}
\usepackage{tabularx}
\usepackage{booktabs} 

\let\labelindent\relax
\usepackage{enumitem}

\usepackage{tikz}

\usepackage{ifthen}
\usepackage{amssymb}
\newboolean{showcomments}
\setboolean{showcomments}{true} 
\ifthenelse{\boolean{showcomments}}
{\newcommand{\nb}[2]{
    \fcolorbox{gray}{yellow}{\bfseries\sffamily\scriptsize#1}
    {$\blacktriangleright$#2$\blacktriangleleft$}
  }
  
}
{\newcommand{\nb}[2]{}
  
}

\usepackage{subcaption}

\newcommand{\eg}{e.g.,~}							
\newcommand{\ie}{i.e.,~}							
\newcommand{\Fig}[1]{Figure~\ref{#1}}  			
\newcommand{\Table}[1]{Table~\ref{#1}}	    

\newcommand{\scorr}{%
  \tikz[baseline=(char.base)]{\node[shape=circle,fill=black!60!green, minimum size=0.5cm] (char) at (0,0) {\vphantom{A}};}
}

\newcommand{\corr}{%
  \tikz[baseline=(char.base)]{\node[shape=circle,fill=black!10!green!50, minimum size=0.5cm, anchor=center] (char) at (0,0) {\vphantom{A}};}
}

\newcommand{\sacorr}{%
 \tikz[baseline=(char.base)]{\node[shape=circle,fill=black!10!red, minimum size=0.5cm, anchor=center] (char) at (0,0) {\vphantom{A}};}
}

\newcommand{\acorr}{%
  \tikz[baseline=(char.base)]{\node[shape=circle,fill=black!10!red!50, minimum size=0.5cm, anchor=center] (char) at (0,0) {\vphantom{A}};}
}

\newcommand{\unrel}{%
  \tikz[baseline=(char.base)]{\node[shape=circle,fill=black!10, minimum size=0.5cm, anchor=center] (char) at (0,0) {\vphantom{A}};}
}

\begin{document}

\sptitle{Theme Article: Special Issue on the Wellbeing for Resilience: Developers Thrive}

\title{Toward Optimal Psychological Functioning in AI-driven Software Engineering Tasks: The SEWELL-CARE Assessment Framework}

\author{Oussama Ben Sghaier}
\affil{Université de Montréal, Canada}

\author{Jean-Sébastien Boudrias}
\affil{Université de Montréal, Canada}

\author{Houari Sahraoui}
\affil{Université de Montréal, Canada}

\markboth{The SEWELL-CARE Assessment Framework}{The SEWELL-CARE Assessment Framework}

\begin{abstract}
In the field of software engineering, there has been a shift towards utilizing various artificial intelligence techniques to address challenges and create innovative tools. These solutions are aimed at enhancing efficiency, automating tasks, and providing valuable support to developers. While the technical aspects are crucial, the well-being and psychology of the individuals performing these tasks are often overlooked. This paper argues that a holistic approach is essential, one that considers the technical, psychological, and social aspects of software engineering tasks.
To address this gap, we introduce SEWELL-CARE, a conceptual framework designed to assess AI-driven software engineering tasks from multiple perspectives, with the goal of customizing the tools to improve the efficiency, well-being, and psychological functioning of developers. By emphasizing both technical and human dimensions, our framework provides a nuanced evaluation that goes beyond traditional technical metrics.
\end{abstract}

\maketitle
 \section{Introduction}

\chapteri{T}he rise of artificial intelligence marks a profound transformation in the role of software developers. A plethora of AI-driven assistants has been developed and evolved to support developers in their tasks \cite{yang2022survey}. These advancements in AI have empowered us to tackle a broader spectrum of software engineering tasks, address more intricate challenges, and elevate the overall degree of automation. This has enabled researchers to propose sophisticated tools endowed with assistive or automation capacities that exhibit heightened accuracy, intelligence, automation, and efficiency. The goal is to enhance the efficiency of these tasks and mitigate prevailing challenges.

However, software engineering problems were predominantly addressed from a technical perspective. 
The conventional approach for evaluating the proposed tools, in software engineering, has been mainly centered on technical metrics \cite{sai2022survey}. 
For instance, performance metrics such as precision, recall, BLEU score, etc. are commonly used as objective measures to assess the performance and efficacy of AI-based tools.

The prevailing tendency to prioritize technical evaluations leads to overlooking the subjective experience, psychology, and cognitive needs of the user.
However, the proposed tools are mainly utilized by humans (\eg developers). 
That is, human aspects constitute an essential dimension that is often underexplored. 
This suggests the inadequacy of purely technical evaluation, as such an assessment remains incomplete, rendering the tools impractical and ineffectual in real-world scenarios. The pivotal role of the user accentuates the need to extend the evaluation beyond mere technical parameters \cite{hazzan2004human}. Consequently, a comprehensive and equitable assessment necessitates the inclusion of human factors associated with software engineering tasks, encompassing elements related to the well-being of the user. This entails considering elements such as user satisfaction, stress levels, commitment, autonomy, well-being, etc. which are paramount to have a complete and fair evaluation.
Incorporating these factors not only contributes to the development of more robust and adapted tools but also fosters a conducive work environment.

Promoting employee well-being has been shown to enhance work engagement \cite{cropanzano2001happy}, support employee autonomy \cite{staw1994employee}, foster organizational citizenship behaviors \cite{lee2002organizational}, and improve productivity \cite{patterson2004organizational}.
Thus, the performance and reliability of the produced software should not be the only criteria that guide the development of AI-based tools. Yet, aspects of human development for the different software engineering actors should also be considered
The trade-off between technical aspects and human factors must be carefully navigated to ensure that the software engineering tools not only enhance efficiency but also support the well-being and optimal psychological functioning of those involved.

Optimal psychological functioning is an area of study that focuses on understanding how individuals become the best that they can be and how to achieve their full potential \cite{chenard2023optimal}. It reflects the paradigm of positive psychology and is concerned with a person’s achievement of maximization in his functioning, whether mental, cognitive, emotional, or social \cite{phan2018importance}.

In this paper, we propose a comprehensive and holistic assessment framework called SEWELL-CARE (\emph{Software Evaluation for WELL-being and optimal psychological functioning in a Context-AwaRe Environment}). 
SEWELL-CARE employs different attributes to accurately characterize AI-driven software engineering tasks.
It also presents the potential outcomes that may have an impact on developers or other tasks. 
This enables a fair evaluation of software engineering tasks that complements technical measures with psychological and human aspects.
Given a specific context, SEWELL-CARE allows the design of adequate AI-based assistants with the aim of maximizing the desired outcomes.
This makes it possible to build complete tools that accurately assist developers in their tasks, ensure their well-being, and allow them to achieve optimal psychological functioning.

We use a real-world scenario to illustrate the usage of SEWELL-CARE. 
We investigate the relation between selected subsets of characteristics and outcomes.
Then, we discuss the optimal task design to maximize the desired outcomes.

The remainder of this paper is organized as follows.
First, we present the proposed assessment framework.
Second, we illustrate the usage of our framework with a real-world scenario. 
Third, we outline some related work.
Finally, we conclude.

\section{SEWELL-CARE: An assessment framework for AI-driven software engineering tasks}
\label{sec:framework}

\begin{figure*}[!htbp]
\centering
\includegraphics[width=1\linewidth]{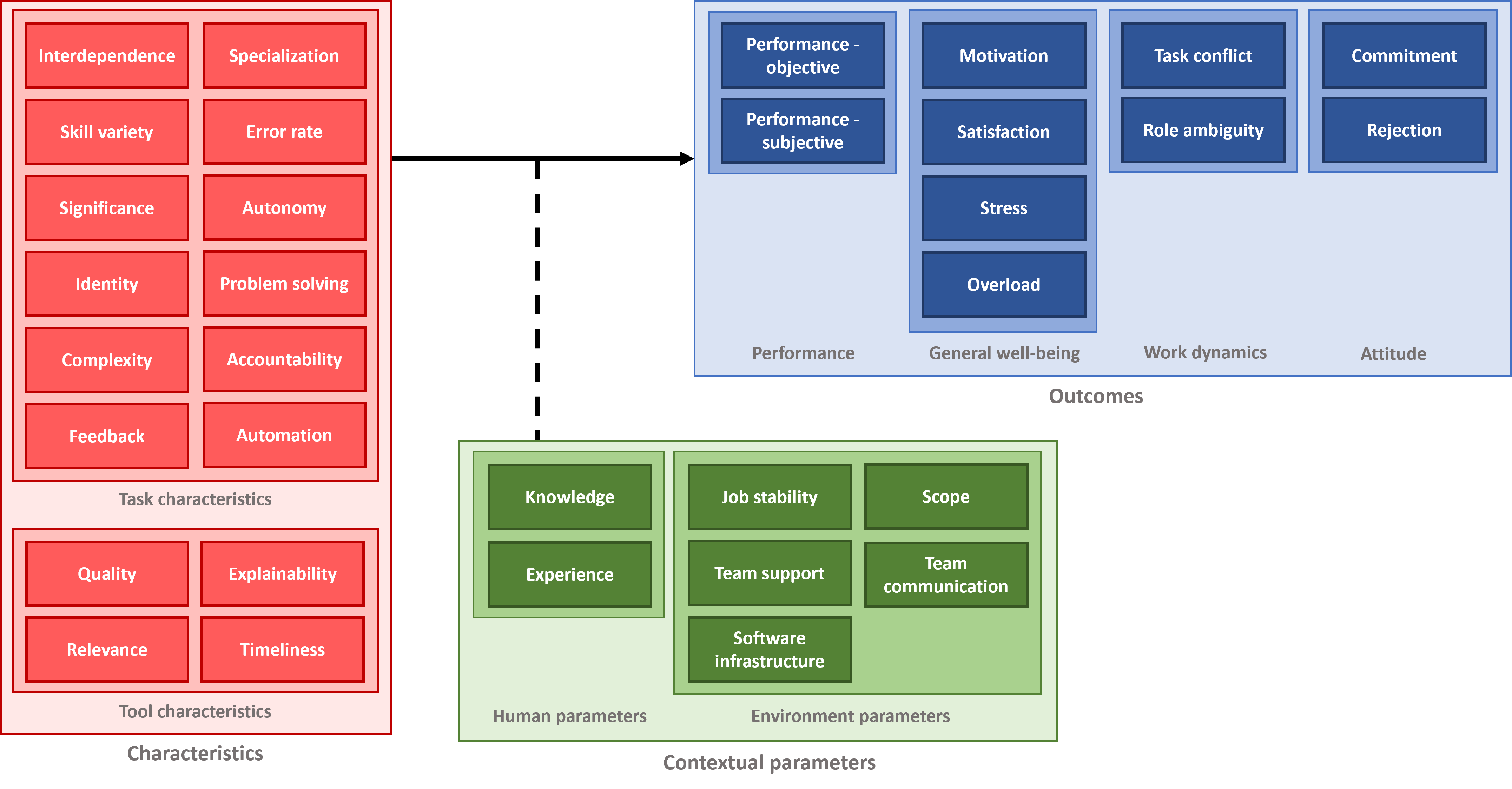}
\caption{SEWELL-CARE: An assessment framework for AI-driven software engineering tasks}
\label{fig:framework}
\end{figure*}

We propose a conceptual framework, named SEWELL-CARE, for assessing AI-driven software engineering tasks. 
In this framework, we consider the impact of the AI-driven task and its associated tool on the well-being and psychological functioning of the developer. This impact is also influenced by factors related to the developer and the environment.
\Fig{fig:framework} presents a schematic representation of SEWELL-CARE.

\begin{table}[!htbp]
  \centering
  \caption{Description of the characteristics of software engineering tasks }
  \label{tab:task_chars}
  \begin{tabularx}{\linewidth}{l X}
    \toprule
    \textbf{Characteristic} & \textbf{Description}\\
    \midrule
    Interdependence & Degree to which a task is dependent on other tasks/other tasks depend on.\\
    Skill variety & Degree to which the task requires a variety of skills to be accomplished.\\
    Significance & Degree to which the task influences people or other tasks.\\
    Identity & Degree to which a task involves a set of activities that identify the task.\\
    Complexity & Degree a task is complex and difficult to perform (\ie requires high-level skills and is   more mentally demanding and challenging).\\
    Feedback & Refers to information, opinions, or evaluations provided by others with the aim of offering insight, guidance, or assessment.\\
    Specialization & Extent to which a task requires specific knowledge and skills.\\
    Error rate & Frequency and percentage of mistakes to which a task is tolerant.\\
    Autonomy & Degree of freedom and independence to accomplish a task (work scheduling, decision making, work methods).\\
    Problem solving & Degree to which a task requires unique solutions and innovative ideas.\\
    Accountability & Extent to which a person is responsible for the task.\\  
    Automation & Degree to which a task is automated or involves the use of assistant/automation tools.\\
    \addlinespace\hline\addlinespace
    Explainability & The extent to which a developer understands the results provided to him by the tools.\\
    Quality & Quality of the results generated by the tools.\\
    Relevance & Relevance of the provided results.\\
    Timeliness & Refers to how often and how quickly is adequate assistance provided.\\    
    \bottomrule
\end{tabularx}
\end{table}

\begin{table*}[!htbp]
  \centering
  \caption{Description of outcomes}
  \label{tab:outcomes}
  \begin{tabularx}{\linewidth}{l X}
    \toprule
    \textbf{Outcome} & \textbf{Description}\\
    \midrule
    Performance–objective & Objective performance is based on measurable and quantifiable data. It involves concrete and factual criteria that are not influenced by personal opinions or emotions.\\
    Performance–subjective & Subjective performance is based on personal opinions, perceptions, or judgments. It relies on the person’s personal feelings, beliefs, and interpretations.\\
    \addlinespace\hline\addlinespace
    Motivation & Refers to individual energization, influencing his initiation, persistence, and enthusiasm in pursuing and completing the task successfully.\\
    Satisfaction & Person’s contentment and fulfillment for the work done in a particular task in accordance with expectations and requirements.\\
    Stress & Refers to the heightened mental or emotional strain experienced by an individual due to perceived pressure, demands, or challenges associated with the task.\\
    Overload & Refers to a condition where an individual is subjected to an excessive amount of work or information surpassing their capacity and potentially leading to difficulties in managing and completing tasks effectively.\\
    \addlinespace\hline\addlinespace
    Task conflict & Refers to the perceived disagreements with other tasks.\\
    Role ambiguity & Refers to the perceived uncertainty or lack of clarity regarding one's task duties, responsibilities, or expectations.\\
    \addlinespace\hline\addlinespace
    Commitment & Refers to dedication, involvement, and responsibility to carry out and complete the task.\\
    Rejection & Implicit or explicit disapproval or refusal to accomplish/be assigned to a task.\\
    \bottomrule
\end{tabularx}
\end{table*}

\begin{table}[!htbp]
  \centering
  \caption{Description of contextual parameters}
  \label{tab:ctxt_chars}
  \begin{tabularx}{\linewidth}{l X}
    \toprule
    \textbf{Parameter} & \textbf{Description}\\
    \midrule
    Experience & Refers to the experience of the person who is accomplishing the task.\\
    Knowledge & Refers to the knowledge and expertise of the person who is doing the task.\\\addlinespace \hline \addlinespace 
    Job stability & Refers to the security and predictability of one's employment within the organization.\\
    Team support & Refers to the level of help, assistance, collaboration, and resources provided by team members.\\
    Software infrastructure & Refers to the foundational software resources and technologies available to support the task, including hardware, code base quality, and infrastructure that can affect task execution.\\
    Scope & Encompasses the comprehensive framework of the task, which includes factors such as project deadlines, client or internal project distinctions, and the overall context within which the task operates.\\
    Team communication & Involves the quality, frequency, and effectiveness of communication  and coordination among team members.\\
    \bottomrule
\end{tabularx}
\end{table}

\paragraph{\textbf{Characteristics}}
The left side of \Fig{fig:framework} serves as a representation of task and tool characteristics. 
These attributes encapsulate features pertaining to software engineering tasks.
Additionally, we include attributes related to AI tools used as assistants for these tasks.
The delineation of a software engineering task in practical scenarios is based on the calibration and instantiation of these properties.
\Table{tab:task_chars} defines these characteristics.

\paragraph{\textbf{Outcomes}}
The right side delineates the potential outcomes associated with software engineering tasks. 
These cover the possible consequences of a task both on developers and on other tasks.
\Table{tab:outcomes} outlines these outcomes.
The outcomes are organized into distinct categories.
The category denoted as \emph{performance} includes technical metrics of both qualitative and quantitative nature.
The \emph{general well-being} category comprises a spectrum of psychological factors that influence the overall state of mental health, happiness, and contentment experienced by individuals when completing a task.
The \emph{work dynamics} category encompasses outcomes that impact other tasks or the work in general. 
Lastly, the \emph{attitude} category covers attributes related to the behavior and demeanor of practitioners with respect to the task under consideration.

\paragraph{\textbf{Contextual parameters}}
The relationship between the characteristics and outcomes is contingent upon a specific context. It depends on the environment as well as the person who is accomplishing the task.
\Table{tab:ctxt_chars} details these parameters.

The contextual parameters could be split into two major groups.
The \emph{human parameters} group refers to the set of attributes that characterize the person doing the task.
The \emph{environment parameters} group includes attributes that characterize the job environment.

In our proposed framework, we defined a three-dimensional space comprising characteristics, contextual parameters, and outcomes. Our primary objective is to investigate the interrelationship between these three dimensions, specifically focusing on the correlation between characteristics and outcomes across various contexts. It is important to note that the analysis of such relationships is inherently intricate due to the vast number of possible permutations when working within the constraints of a three-dimensional space.

To facilitate a more manageable analysis, we propose to practitioners a strategy that involves partitioning the investigation into discrete components, by considering specific use cases. That is, we need to establish the context by explicitly setting the human and environment parameters. Within a particular context or setting, we are equipped to conduct a comprehensive correlation analysis between the characteristics of a given task and its potential outcomes. This analysis allows us to pinpoint the characteristics that significantly impact each outcome and, crucially, to assess the direction (\ie positive or negative) and strength of these correlations. Consequently, by gaining a deeper understanding of these relationships, we can strategically optimize the design of the associated tasks to achieve the desired outcomes effectively and efficiently.

\section{Illustrative example}
\label{sec:evaluation}

To demonstrate the practical applicability of \mbox{SEWELL-CARE}, we employ a real-world scenario involving the code review task.
Within this scenario, two Python developers, named \emph{Vinh} and \emph{Bakhta}, are employed by the same company. They work on a platform for financial data analysis. The two developers are regularly involved in reviewing the code of their fellows.
\emph{Bakhta} is an experimented developer, with several years of experience working for other companies. She has been regularly involved in the process of code review.
\emph{Vinh} is an entry-level Python developer, representing a more junior role. He hasn't previous experience in performing code review.



Each case within the study represents a unique contextual setting.
While these cases share most of the environment parameters, they represent two distinct profiles.
We instantiate the contextual parameters for each use case as detailed in \Table{tab:ctxt_eval}.

\begin{table}[!htbp]
  \centering
  \caption{Contextual parameters assessment for the case study }
  \label{tab:ctxt_eval}
  \begin{tabular}{l  l  c  c}
    \toprule
    & \textbf{Parameter} & \textbf{Vinh} & \textbf{Bakhta}\\
    \midrule
    \multirow{2}{*}{Human} & Knowledge & 1 & 5 \\
        & Experience & 1 & 4 \\\midrule
    \multirow{4}{*}{Environment} & Job stability & 2 & 4 \\
        & Team support & 4 & 4 \\
        & Software infrastructure & 2 & 2 \\
        & Team communication & 4 & 4  \\        
    \bottomrule
  \end{tabular}
\end{table}

To set the context, we assign a value to each parameter in a \emph{1-5 rating scale}.
Notably, both developers operate within a common environment, leading to identical values for the corresponding environment parameters except for \emph{job stability}.
For the human parameters, \emph{Bakhta} is a more knowledgeable and experienced developer in comparison to \emph{Vinh}.

\subsection{Analysis}

For each case, we conduct an in-depth investigation into the influence of varying levels of these task characteristics (low, medium, high) on the respective outcomes, employing a correlation matrix. Our analysis is guided by various research contributions on motivation and tolerance to complexity, \eg \cite{ramachandran1999science, csikszentmihalhi2020finding}. 

To streamline the analysis, we focus on a subset of characteristics and outcomes.
Specifically, we chose to examine the characteristics of \emph{significance}, \emph{automation}, \emph{complexity}, and \emph{problem solving}. In the case of code review, AI-driven tools can offer different levels of assistance, ranging from indicating the presence of a potential issue in the code to suggesting code changes, including the problem localization in the code, and composing review comments. Each level of assistance determines the levels of the considered characteristics.
For the outcomes, we select \emph{performance-objective}, \emph{stress}, \emph{motivation}, and \emph{rejection}.

We investigate the relationship between the characteristics and outcomes. \Table{tab:corr} reports the correlations for each case. 
These correlations offer valuable insights into how the characteristics differently impact various outcomes of software engineering tasks according to the given context.

\newcolumntype{C}[1]{>{\centering\arraybackslash}m{#1}}
\begin{table*}[!htbp]
  \centering
  \caption{Correlation between a subset of characteristics and outcomes}
  \label{tab:corr}
  \begin{tabular}{l|C{2cm}|C{2cm}|C{2cm}|C{2cm}}
    \toprule

    & \textbf{Performance-Objective} & \textbf{Stress}  & \textbf{Motivation} & \textbf{Rejection}\\
    \midrule
    \rowcolor{black!10} \multicolumn{5}{c}{\textcolor{black}{\textbf{Bakhta}}}\\\midrule
    
    \textbf{Significance}&\scorr&\acorr&\scorr&\sacorr\\\midrule
    
    \textbf{Automation}&\corr&\corr&\acorr&\unrel\\\midrule
    
    \textbf{Complexity} & \unrel & \corr & \corr & \corr\\\midrule
    
    \textbf{Problem solving} & \unrel & \corr & \scorr & \acorr\\\midrule

    \rowcolor{black!10} \multicolumn{5}{c}{\textcolor{black}{\textbf{Vinh}}}\\\midrule
    
    \textbf{Significance}&\acorr&\scorr&\acorr&\corr\\\midrule
    
    \textbf{Automation}&\scorr&\sacorr&\scorr&\sacorr\\\midrule
    
    \textbf{Complexity} & \sacorr & \scorr & \sacorr & \scorr\\\midrule
    
    \textbf{Problem solving} & \acorr & \scorr & \sacorr & \corr\\\bottomrule

    \end{tabular}
    \begin{tabular}{|c@{\hspace{3pt}}c|c@{\hspace{3pt}}c|c@{\hspace{3pt}}c|c@{\hspace{3pt}}c|c@{\hspace{3pt}}c|}
    \midrule
    \sacorr & strongly negatively correlated &
    \acorr & negatively correlated &
    \unrel & unrelated &
    \corr & correlated &
    \scorr & strongly correlated \\
    \midrule
    \end{tabular}    
\end{table*}

\paragraph{\textbf{Significance}}
Task significance carries varying weight for experienced and novice developers.
For the experienced developer, \emph{Bakhta}, a task's significance resonates as a hallmark of importance, bolstering her performance and motivation. The greater the significance, the more it piques her interest, alleviates stress, and fosters a willingness to accept the task.

Conversely, for the novice developer, \emph{Vinh}, the dynamics shift. The weight of task significance exerts a different influence. In his case, increased task significance may, paradoxically, hinder his performance and motivation. As a less experienced developer, elevated significance can increase stress levels, ultimately leading to task rejection in cases of heightened importance and criticality.

\paragraph{\textbf{Automation}}
In the context of automation, the impact on \emph{Bakhta}'s performance is notable: the higher the level of task automation, the more proficient she becomes. However, this increase in performance is accompanied by reduced motivation, as heightened automation tends to render tasks less challenging and more repetitive, thereby diminishing their intrinsic appeal. Paradoxically, a substantial increase in task automation may also lead to elevated stress for \emph{Bakhta}, as it diminishes her fine-grained control over the task.

Conversely, for \emph{Vinh}, the consequences of automation take a different turn. Elevated task automation enhances both his performance and motivation, offering valuable assistance and simplifying tasks. This reduction in complexity and the assistance provided by automation serve to alleviate stress, making him more inclined to engage in and accept this kind of task facilitated by automation tools, as it aligns well with his role as a junior developer.

\paragraph{\textbf{Complexity}}
\emph{Bakhta} demonstrates consistent performance across tasks of varying complexity, handling both easy and complex tasks with equal proficiency. However, a distinctive pattern emerges in her emotional response: for more challenging tasks, she experiences increased stress yet heightened motivation. Her extensive experience in the field renders complex tasks more attractive, reflecting her keen interest in tackling them.

In contrast, \emph{Vinh} lacks the extensive experience of his counterpart. Thus, complex tasks tend to diminish his motivation, heighten his stress levels, and lead to decreased performance. Consequently, he tends towards rejecting complex tasks, given that they present greater challenges that could surpass his current skill set and experience level.

\paragraph{\textbf{Problem solving}}
\emph{Bakhta} prefers tasks that demand substantial problem-solving and creative prowess. Engaging in such tasks may induce a mild level of stress in her, but it significantly heightens her motivation to successfully complete them.

\emph{Vinh} tends to avoid assignments that involve high levels of problem solving and creativity, primarily due to his comparatively lower experience and skills. These tasks tend to evoke a high degree of stress in him and are accompanied by reduced motivation.

\section{Related work}
\label{sec:literature}

Many works attempted to address work design and work psychology in software engineering.

In \cite{morgeson2006work}, the authors introduce a comprehensive measure known as the Work Design Questionnaire (WDQ).  This WDQ was devised to define and characterize a job and to measure these characteristics. The identified work characteristics were categorized into three groups: motivational, social, and contextual characteristics. Each of these work characteristics was associated with a distinct set of items, to define and facilitate their measurement through a straightforward 5-point scaling system. The authors extend this work in \cite{humphrey2007integrating}, delving into the impact of work characteristics on outcomes. This exploration was undertaken through a series of hypothesized relationships, that were validated with a user study.

In \cite{forsgren2021space}, the authors introduce \emph{SPACE}, a multi-dimensional framework for developer productivity. It captures different dimensions of productivity including satisfaction, performance, activity, communication, and efficiency. This enables an understanding of developer productivity and thus ensures the efficient development of software systems and the well-being of developers.

In \cite{greiler2022actionable}, the authors propose the \emph{DX Framework}, an actionable conceptual framework for understanding and improving developer experience to enable more productive and effective work environments.
The framework elucidates factors that affect developer experience and characteristics that influence their respective importance to individual developers. It also outlines barriers that hinder the developer experience and presents strategies employed by individuals and teams to improve the developer experience. This conceptual framework is built to understand and guide developer experience improvements.

While these contributions present practical and insightful perspectives, a notable void remains unaddressed. None of the extant endeavors has delved into software engineering tasks, as these works predominantly operate at the macroscopic job level. Moreover, some of these contributions selectively focalize their attention on specific outcomes. The focus of these works primarily centers on developers and the job in a broader sense, without due consideration for the integral role of AI tools in the execution of these tasks.

\section{Conclusion}
\label{sec:conclusion}
In this paper, we introduce SEWELL-CARE, a novel conceptual framework tailored for the evaluation of AI-driven software engineering tasks. This framework serves to characterize AI-driven software engineering tasks within diverse contextual settings and to comprehensively encapsulate their potential outcomes, encompassing psychological dimensions.
This comprehensive framework enables practitioners to strategically tailor task designs, to achieve the desired outcomes. The overarching aim is to foster the design of well-balanced tasks, optimizing their efficiency and concurrently ensuring the well-being of the developers involved in their execution.

The proposed framework currently adopts a descriptive nature, elucidating the relationships between task characteristics and their associated outcomes. As part of future work, we aspire to transition from a descriptive stance to a predictive one. To this end, we plan to concretize these relationships by developing predictive models that quantitatively gauge outcomes based on the task characterizations within specific contextual settings. This transition will empower us to forecast and anticipate outcomes, thereby enhancing the framework's utility and its capacity to practically guide optimized task designs.

\bibliographystyle{IEEEtran}
\bibliography{main}


\begin{IEEEbiography}{Oussama Ben Sghaier}{\,} is a Ph.D. student at the Department of Computer Science and Operations Research of the Université de Montréal at Montréal, Québec, Canada. His research interests include artificial intelligence for software engineering, natural language processing, and model-driven engineering. Contact him at oussama.ben.sghaier@umontreal.ca.
\end{IEEEbiography}

\begin{IEEEbiography}{Jean-Sébastien Boudrias} {\,} is a professor at the Department of Psychology of the Université de Montréal at Montréal, Québec, Canada. His research interests include employee empowerment, personality, and psychological health at work. Contact him at jean-sebastien.boudrias@umontreal.ca.
\end{IEEEbiography}

\begin{IEEEbiography}{Houari Sahraoui}{\,} is a professor at the Department of Computer Science and Operations Research of the Université de Montréal at Montréal, Québec, Canada. His research interests include artificial intelligence techniques applied to software engineering, search-based software engineering, and model-driven engineering. Contact him at sahraouh@iro.umontreal.ca.
\end{IEEEbiography}

\end{document}